\documentclass[11pt]{elsarticle}

\pdfoutput=1

\makeatletter
\def\ps@pprintTitle{%
  \let\@oddhead\@empty
  \let\@evenhead\@empty
  \let\@oddfoot\@empty
  \let\@evenfoot\@oddfoot
}
\makeatother

\usepackage{url}
\usepackage{breakurl}
\usepackage[breaklinks,
            colorlinks = true,
            linkcolor = blue,
            urlcolor  = blue,
            citecolor = blue,
            anchorcolor = blue]{hyperref}

\usepackage{lineno}

\usepackage{graphicx}
\usepackage[margin=1.25in]{geometry}
\usepackage[usenames,dvipsnames]{color}

\newcommand\acp{\begin{center}
\rule[-0.2in]{\hsize}{0.01in}\\\rule{\hsize}{0.01in}\\
\vskip 0.1in Submitted to the  Proceedings\\ 
of the African Conference on Fundamental and Applied Physics
    \vskip 0.05in
    {\it Second Edition, ACP2021, March 7--11, 2022 --- Virtual Event}\\
\rule{\hsize}{0.01in}\\\rule[+0.2in]{\hsize}{0.01in} \\
\end{center}}

\usepackage[firstpage=true]{background}
\backgroundsetup{contents={\parbox{6.5in}{\acp}}, scale=1,placement=top,opacity=1,color=black,position={3.25in,1.2in}}

\usepackage{fancyhdr}
\fancypagestyle{plain}{%
  \fancyhf{}%
  \fancyhead[C]{}
  \fancyfoot[C]{\thepage}
}

\fancypagestyle{empty}{%
  \fancyhf{}%
  \fancyhead[C]{{\it ACP2021, March 7--11, 2022 --- Virtual Edition}}
  \fancyfoot[C]{\thepage}
}
\begin{document}

\begin{frontmatter}


\title{Vacuum stability of the scalar potential in the compact 341 model}

\author[]{Meriem Djouala\corref{cor1}}
\ead{djoualameriem@gmail.com}
\author[]{Noureddine Mebarki}

\cortext[cor1]{Corresponding Author}

\address{Laboratoire de physique math\'{e}matique et subatomique, university of Fr\`{e}res Mentouri-Constantine 1-Algeria.}

\begin{abstract}
\noindent 
By applying the concepts of copositivity and using the gauge orbit spaces on the scalar potential, we derive analytic necessary and sufficient conditions which guarantee the boundedness of the scalar potential in all the directions in the field space at the tree level in the
context of the compact 341 model.
\end{abstract}

\begin{keyword}
Beyond the Standard Model \sep the comapct 341 model \sep vacuum stability
\end{keyword}

\end{frontmatter}

\section{Introduction}
\label{sec:intro}
\noindent
~~~The Standard Model (SM) failed to address the answer of many outstanding questions such as neutrinos oscillations, dark matter,etc.; thus, going beyond it becomes mandatory. The SM can be extended in several different ways, by adding new fermions fields \cite{Djouala2,Djouala3}, augmenting the scalar sector to more than Higgs field \cite{Garv,J} and enlarging the local gauge group \cite{Dias}. Among the theories beyond the Standard Model, we are interested in the second extension of the electroweak gauge group of SM, the model based on the gauge group $SU(3)_{C}\otimes SU(4)_{L}\otimes U(1)_{X}$ (341 model just for short) where $C$ refers the color charge, $L$ is the left handed chirality while $X$ is a new quantum number associated to the group $U(1)_X$.
This model predicts the existence of new particles included the exotic leptons and quarks, new gauge and scalar bosons.\\
\indent To specify the allowed regions of the scalar parameters one needs to derive the theoretical constraints besides the experiment limits. The vacuum stability condition determines the first set of the theoretical constraints; in our work, we derive the necessary conditions that ensure the boundedness of the scalar potential from below in any direction in the field space in the compact 341 model \cite{Djouala} by using the copositivity criteria and the gauge orbit space.\\
\indent This paper is organized as follows. In Section \ref{sec:341}, we present the fermion content of the compact 341 model. In Section \ref{sec:res}, we discuss the vacuum stability conditions using the copositivity criteria and the gauge orbit space. Finally, in Section, \ref{sec:Con} we draw our conclusions. 
\section{The compact 341 model}
\label{sec:341}
\noindent
\indent In the compact 341 model, the electric charges of new particles are defined through the following electric charge operator
\begin{small}
	\begin{equation}
	Q=\frac{1}{2}\bigg(\lambda_{3}-\frac{1}{\sqrt{3}}\lambda_{8}-\frac{4}{\sqrt{6}} \lambda_{15}\bigg)+X,
	\end{equation}
\end{small}
where $\lambda_{m=3,8,15}$ are the diagonal matrices of the group $SU(4)$. The fermion content in this model is
\begin{equation}
\psi_{aL}\equiv\left(
\begin{array}{ccc}
\nu_{a} \\
l_{a} \\
\nu_{a}^{c}\\
l_{a}^{c} \\
\end{array}
\right)_{(1,4,0)},
Q_{1L}\equiv\left(
\begin{array}{ccc}
u_{1} \\
d_{1} \\
U_{1}\\
J_{1} \\
\end{array}
\right)_{(3,4,\frac{2}{3})},Q_{iL}\equiv\left(
\begin{array}{ccc}
d_{i} \\
u_{i} \\
D_{i}\\
J_{i} \\
\end{array}
\right)_{(1,\bar{4},-\frac{1}{3})}
\end{equation}
where $a\equiv e,\mu,\tau$. $Q_{1L}$ is the first left handed quark generation and $Q_{iL(i=2,3)}$ represent the second and third families of left handed quarks.\\
\indent To cancel the gauge anomalies in the compact 341 model, the quark generations should belong to different representations, one of the quark families $Q_{1L}$ with the three left handed lepton generations $\psi_{aL}$ arrange in the fundamental representation whereas the other two families $Q_{iL(i=2,3)}$ are in the conjugate fundamental representation (or vice versa).\\
\indent Notice that the right-handed quarks transform as singlets under $SU(4)_{L}\otimes U(1)_{X}$, while, the right handed leptons belong to the left handed quadruplets.\\
\indent The most general scalar potential in the compact 341 model is given by
\begin{small}
	\begin{eqnarray}\label{eq:mi}
	V(\eta,\rho,\chi)&=&\mu_{\eta}^{2}\eta^{\dag}\eta+\mu_{\rho}^{2}\rho^{\dag}\rho+\mu_{\chi}^{2}\chi^{\dag}\chi+\lambda_{1}(\eta^{\dag}\eta)^{2}+\lambda_{2}(\rho^{\dag}\rho)^{2}+\lambda_{3}(\chi^{\dag}\chi)^{2}
	\nonumber\\&+&\lambda_{4}(\eta^{\dag}\eta)(\rho^{\dag}\rho)+\lambda_{5}(\eta^{\dag}\eta)(\chi^{\dag}\chi)+\lambda_{6}(\rho^{\dag}\rho)(\chi^{\dag}\chi)+\lambda_{7}\nonumber
	(\rho^{\dag}\eta)(\eta^{\dag}\rho)
	\\&+&\lambda_{8}(\chi^{\dag}\eta)(\eta^{\dag}\chi)+\lambda_{9}(\rho^{\dag}\chi)(\chi^{\dag}\rho),
	\end{eqnarray}
\end{small}
where $\mu^{2}_{\mu,\rho,\chi}$ are the mass dimension parameters, $\lambda_{j} (j=1...9)$ are dimensionless coupling constants and $\eta$, $\rho$ and $\chi$ are the scalar fields that are given by
\begin{equation}
\rho\equiv\left(
\begin{array}{ccc}
\rho^+_1 \\
\rho^0 \\
\rho^+_2\\
\rho^{++}\\
\end{array}
\right)_{(1,4,1)},
\chi\equiv\left(
\begin{array}{ccc}
\chi^-_1\\
\chi^{--}\\
\chi^-_2\\
\chi^0\\
\end{array}
\right)_{(3,4,-1)},
\eta\equiv\left(
\begin{array}{ccc}
\eta^0_1\\
\eta_1^-\\
\eta^0_2\\
\eta^+_2
\end{array}
\right)_{(1,4,0)}
\end{equation}
where the VEVs structure is represented as
\begin{equation}
\langle \rho \rangle=\frac{1}{\sqrt{2}}\left(
\begin{array}{ccc}
0 \\
v_\rho \\
0\\
0\\
\end{array}
\right),
\langle \chi \rangle=\frac{1}{\sqrt{2}}\left(
\begin{array}{ccc}
0\\
0\\
0\\
v_\chi\\
\end{array}
\right),
\langle \eta \rangle=\frac{1}{\sqrt{2}}\left(
\begin{array}{ccc}
0\\
0\\
v_\eta\\
0
\end{array}
\right)
\end{equation}
The VEV $v_\chi$ is responsible for the first step in breaking the 341 symmetry to 331, whereas $v_\eta$ breaks the
331 symmetry to 321 (the SM), and $v_\rho$ breaks the SM to $U(1)_{EM}$. Thus one can write $v_\chi>v_\eta>v_\rho$.\\
\indent The conditions for the minimum of the scalar potential (\ref{eq:mi}) are
\begin{eqnarray}\label{eq:the}
\mu_{\eta}^{2}+\lambda_{1}\upsilon_{\eta}^{2}+\frac{1}{2}\lambda_{4}\upsilon_{\rho}^{2}+\frac{1}{2}\lambda_{5}\upsilon_{\chi}^{2}=0,\\
\mu_{\rho}^{2}+\lambda_{2}\upsilon_{\rho}^{2}+\frac{1}{2}\lambda_{4}\upsilon_{\eta}^{2}+\frac{1}{2}\lambda_{6}\upsilon_{\chi}^{2}=0,\label{eq:the2}\\
\mu_{\chi}^{2}+\lambda_{3}\upsilon_{\chi}^{2}+\frac{1}{2}\lambda_{5}\upsilon_{\eta}^{2}+\frac{1}{2}\lambda_{6}\upsilon_{\rho}^{2}=0 \label{eq:the3}.
\end{eqnarray}
\section{The vacuum stability}
\label{sec:res}
\indent To ensure the boundedness of the scalar potential in the compact 341 model from below in any direction in the field space we derive analytic necessary and sufficient conditions of the vacuum stability by using the concepts of copositivity and gauge orbit spaces.\\
\indent In the limit of large field
values, it is enough to work with the quartic
terms. We ignore other terms with dimension $d<4$ since they are negligible in comparison with the quartic couplings of the scalar potential, thus $V^{4}(\eta,\rho,\chi)$ takes the following form
\begin{eqnarray}\label{eq:mimzy}
V^{4}(\eta,\rho,\chi)=&\lambda_{1}(\eta^{\dag}\eta)^{2}+\lambda_{2}(\rho^{\dag}\rho)^{2}+\lambda_{3}(\chi^{\dag}\chi)^{2}
+\lambda_{4}(\eta^{\dag}\eta)(\rho^{\dag}\rho)+\lambda_{5}(\eta^{\dag}\eta)(\chi^{\dag}\chi)\nonumber\\&+\lambda_{6}(\rho^{\dag}\rho)(\chi^{\dag}\chi)+\lambda_{7}(\rho^{\dag}\eta)(\eta^{\dag}\rho)+\lambda_{8}(\chi^{\dag}\eta)(\eta^{\dag}\chi)+\lambda_{9}(\rho^{\dag}\chi)(\chi^{\dag}\rho),
\end{eqnarray}
We have three scalar field directions; thus, we define a parametrization as follows \cite{Djouala}
\begin{eqnarray}
r^2&\equiv& \eta^\dag\eta+\rho^\dag\rho+\chi^\dag\chi,\nonumber\\
\eta^\dag\eta&\equiv& r^2\cos^2\theta \sin^2\phi,\nonumber\\
\rho^\dag\rho&\equiv& r^2\sin^2\theta \sin^2\phi,\nonumber\\
\chi^\dag\chi&\equiv& r^2\cos^2\phi,\nonumber\\
\frac{\eta^\dag\rho}{|\eta||\rho|}&\equiv &\xi_{1}e^{i\psi_{1}},\nonumber \\
\frac{\eta^\dag\chi}{|\eta||\chi|}&\equiv& \xi_{2}e^{i\psi_{2}}, \nonumber\\
\frac{\rho^\dag\chi}{|\rho||\chi|}&\equiv& \xi_{3}e^{i\psi_{3}},
\end{eqnarray}
where the radius $r$ scans the domain [0,$\infty$[, $\theta \in$ [0,2$\pi$], $\phi\in$[0,$\frac{\pi}{2}$] and $\xi_{i=(1,2,3)}\in$ [0,1] \cite{Djouala}.\\
Inserting this parametrization in Eq.(\ref{eq:mimzy}), we get expressions that have the following form \cite{Djouala}
\begin{eqnarray}\label{eq:mimzy55}
f(\chi)=a\chi^2+b(1-\chi)^2+c\chi(1-\chi),
\end{eqnarray}
The conditions for copositivity are given below \cite{Djouala}:
\begin{eqnarray}\label{eq:mimz55}
a>0,~~ b>0~~c+2\sqrt{ab}>0.
\end{eqnarray}
\indent Notice that the quartic part of the scalar potential $(\ref{eq:mimzy})$ is bounded from below only if it satisfies the copositivity conditions (\ref{eq:mimz55}). Thus, by
applying this criteria to our results, we find \cite{Djouala}:
\begin{eqnarray}
\lambda_{1}&>&0,~~\lambda_{2}>0~~ \lambda_{3}>0, \nonumber\\ \lambda_{4}&+&2\sqrt{\lambda_{1}\lambda_{2}}>0\nonumber.\\ \lambda_{4}&+&\lambda_{7}+2\sqrt{\lambda_{1}\lambda_{2}}>0 \label{eq:az}.
\end{eqnarray}
The conditions (\ref{eq:az}) ensure the stability of the scalar potential in the compact 341 model, together with the positivity of the Hessian matrix and the perturbative unitarity conditions we get the set of the theoretical constraints on the scalar couplings \cite{Djouala}.
\section{Conclusions}
\label{sec:Con}
throughout this paper, we have studied the vacuum stability constraint in the compact 341 model which determines the necessary and sufficient
conditions for the boundedness of the scalar potential from below in any direction in the field space by using the copositivity criteria and gauge orbit space. Together with the other theoretical constraints we can determine the allowed and compatible regions of the scalar parameters space \cite{Djouala}.
\section*{Acknowledgments}
We would like to thank Garv Chauhan for fruitful discussions. We are very grateful to the Algerian
ministry of higher education and scientific research and DGRSDT for the financial
support.

\end{document}